\documentclass[preprint,tightenlines,aps,floats]{revtex4}

\usepackage{epsf}
\usepackage{epsfig}
\usepackage{graphicx}
\usepackage[dvips]{color}
\usepackage{amsmath,amssymb,bbm,mathrsfs}
\usepackage{axodraw}

\newcommand{\tpi}{\tilde{\pi}}
\newcommand{\trho}{\tilde{\rho}}

\newcommand{\notD}{\hbox{{$D$}\kern-.60em\hbox{/}}}
\newcommand{\notE}{\ \hbox{{$E$}\kern-.60em\hbox{/}}}
\newcommand{\notp}{\ \hbox{{$p$}\kern-.43em\hbox{/}}}

\newcommand{\Del}{\Delta}

\def\Cred#1{{\textcolor{black}{#1}}}
\def\Cblue#1{{\textcolor{black}{#1}}}
\def\Cgreen#1{{\textcolor{black}{#1}}}

\includegraphics{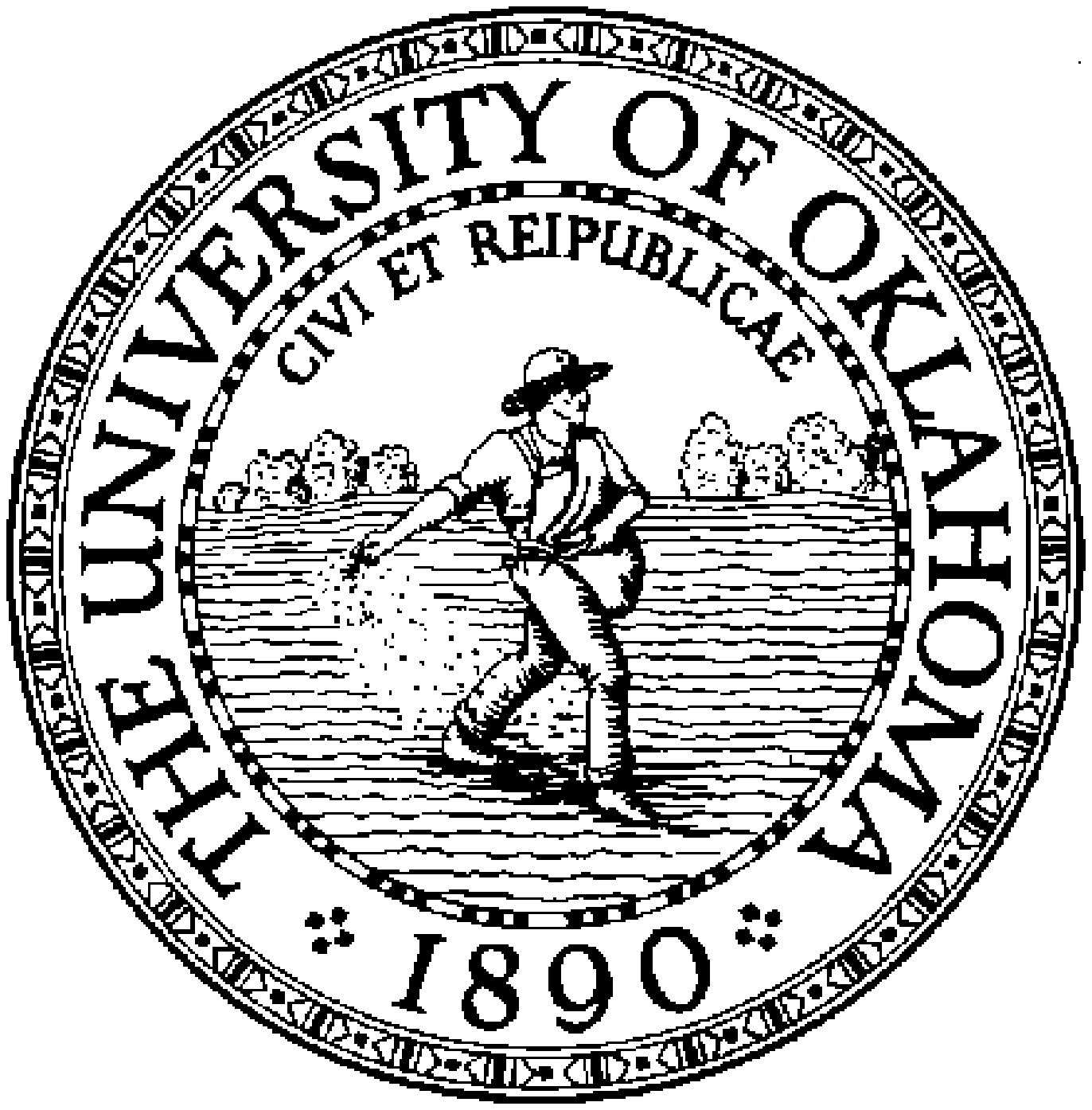}

\preprint{\font\fortssbx=cmssbx10 scaled \magstep2
\hbox to \hsize{
\hskip1.2in %\raise.1in
\hbox{\fortssbx The University of Oklahoma}
\hskip0.2in $\vcenter{
                      \hbox{\bf arXiv: [hep-ph]}
                      \hbox{\bf OSU-HEP-11-2}
                      \hbox{\bf OU-HEP-110122}
                      \hbox{May 2011}}$ }
}

\begin{document}

%-----------------------------------
%   Title
%-----------------------------------
\title{\vspace*{0.7in}
Searching for Colorons at the Large Hadron Collider}

%-----------------------------------
%   Authors
%-----------------------------------
\author{
Joshua Sayre$^a$\footnote{Email address: sayre@physics.ou.edu},
Duane A. Dicus$^b$\footnote{Email address: dicus@physics.utexas.edu},
Chung Kao$^a$\footnote{Email address: kao@physics.ou.edu},
S. Nandi$^c$\footnote{Email address: s.nandi@okstate.edu}}

%-----------------------------------
% Address
%-----------------------------------
\affiliation{
$^a$Homer L. Dodge Department of Physics and Astronomy
and Oklahoma Center for High Energy Physics,
University of Oklahoma,
Norman, OK 73019, USA \\
$^b$Center for Particles and Fields and Texas Cosmology Center,
University of Texas, Austin, TX 78712, USA \\
$^c$Department of Physics and Oklahoma Center for High Energy Physics,
Oklahoma State University, Stillwater, OK 74078, USA
\vspace*{.4in}}

\date{\today}

%-----------------------------------
%   Abstract
%-----------------------------------
\begin{abstract}

We investigate the prospects for the discovery of massive color-octet 
vector bosons at the CERN Large Hadron Collider with $\sqrt{s} = 14$ TeV.
A phenomenological Lagrangian is adopted to evaluate the cross section
of a pair of colored vector bosons (colorons, $\tilde{\rho}$) decaying into
four colored scalar resonances (hyper-pions, $\tilde{\pi}$),
which then decay into eight gluons.
We include the dominant physics background from the production of
$8g,7g1q, 6g2q$, and $5g3q$, 
and determine the masses of $\tilde{\pi}$ and $\tilde{\rho}$ where
discovery is possible.  For example, we find that a 5$\sigma$ signal
can be established for $M_{\tilde{\pi}} \alt 495$ GeV 
($M_{\tilde{\rho}} \alt 1650$ GeV).
More generally we give the reach of this process for a selection of
possible cuts and integrated luminosities.

\end{abstract}

%

%12.60.-i Models beyond the standard model
%12.60.Rc Composite models
%13.85.Ni Inclusive production with identified hadrons
%14.40.Rt Exotic mesons
%14.70.Pw Other gauge bosons

%-----------------------------------------------------------------------
% Make Title Page
%-----------------------------------------------------------------------
\maketitle

%=======================================================================
% BEGIN MAIN TEXT
%=======================================================================
\newpage

\color{black}

%-----------------------------------------------------------------------
% 1 Introduction
%-----------------------------------------------------------------------
\section{Introduction}
With the LHC beginning to accumulate data, we look forward to a new
era of high-energy physics \Cblue{where} we explore multi-TeV energy scales.
In addition to the search for the Higgs boson to complete the
Standard Model, many new physics scenarios have been considered as
potential discoveries. Often, the discovery potential provided by the
LHC's unprecedented collision energies is somewhat mitigated by the
prevalence of jets from Standard Model processes. New physics
which proceeds through weak interactions, such as Higgs production,
must be carefully separated from large, strong-force produced
backgrounds via judicious selection cuts.

It is also possible that new physics will manifest itself through
the strong force. If \Cred{new strongly interacting colored}
particles exist at TeV scales, they will be discovered through
decays into jets. One generic possibility is a massive vector boson
in the color-octet representation. Such a particle has been dubbed a
coloron and several theories of physics
beyond the Standard Model give rise to colorons. These include
Kaluza-Klein excitations of the gluon in extra-dimensional
models~\cite{Dicus:2000hm,Cullen:2000ef}, new states in top-color assisted
technicolor~\cite{Hill:1991at}, as well as models where the standard
color group is a remnant of broken $SU(3) \times
SU(3)$~\cite{Dicus:1994sw}. In Ref.~\cite{Kilic:2008pm}, Kilic, Okui, 
and Sundrum showed how colorons, as well as a scalar octet,
can emerge as the low energy states of an effective theory arising
from a simple model of new, strongly interacting fermions. 
In this paper we follow their analysis, as well as the subsequent
treatment found in Ref.~\cite{Kilic:2008ub}.

Briefly, we suppose that there exists a new, strongly coupled
force, termed hypercolor, which becomes confining at higher
energies than the QCD strong force. Fermions which carry
hypercolor will form bound states which are hypercolor singlets but
which may carry standard model charges. In particular, if these
``hyperquarks" are also triplets of QCD then the lightest bound
states will be color octets. Analogously to the breaking of chiral
symmetry in the standard model, this model will produce relatively 
light hyper-pions as pseudo-Goldstone bosons, as well as colorons 
as bound states in the color octet representation.

%\section{The LHC at 7 TeV}

%The results above are based on the assumption 
This work assumes
that the LHC will achieve its design center of mass energy of 14 TeV and accumulate data at that
energy, and with high luminosity, over several years. 
%Currently the LHC is finishing up a run at a nominal energy of 7 TeV. 
In a separate paper, we have
analyzed this coloron
model for an energy of 7 TeV with an assumed $1~\text{fb}^{-1}$ \Cblue{or 5 fb$^{-1}$} of total integrated luminosity~\cite{Dicus:2010bm}.
We found that there is a chance of early detection
from this \Cblue{early} data set for hyper-pions up to \Cblue{a mass of} $250$ GeV.

%These colorons, and scalar octets can have an interesting
%phenomenology. Naively one might think that light octets are severely
%constrained by dijet searches at the Tevatron. However, in this model
%the hyper-pions couple sufficiently weakly to gluons to leave an
%interesting parameter range and hyper-rhos have only a small branching
%fraction to decay into two quarks or two gluons. Rather, the hyper-rho
%decays predominantly to hyper-pions, which then each decay to a gluon
%pair. Thus the dominant signal for resonant production of the
%hyper-rho is a 4-jet decay chain.

%On the other hand, there are several processes which pair produce
%hyper-pions without a resonant rho. Combined with the loss of jet
%resolution during showering, hadronization, and reconstruction, this
%may make the initial rho resonance difficult to establish. To better
%determine it we also consider the pair production of hyper-rhos,
%leading to an eight-jet signal.

\Cblue{In Section II we give some details of the explicit model we use.
The colorons and octet scalars have an interesting phenomenology which we describe in Sec. III.
In particular, we give some results for resonant $\tilde{\rho}$ production branching to two $\tilde{\pi}$ which decay into four jets.
However, we argue that a more promising channel for coloron detection is coloron pair production followed
by decay to four $\tilde{\pi}$ and then to eight jets.
We show the results for this signal in Sec. IV and for the eight jet background in Sec. V.
In Sec. VI we combine the signal and background to find a range of values of $M_{\tilde{\rho}}$ ($M_{\tilde{\pi}}$)
where the $\tilde{\rho}$ could be discovered.  In Sec. VII we summarize and conclude.}

%-----------------------------------------------------------------------
% 2 A model with colored vector bosons and scalars
%-----------------------------------------------------------------------
\section{A model with colored vector bosons and scalars}

As in Ref.~\cite{Kilic:2008pm}, we assume there is a new
$SU(N)$ gauge group, hypercolor, acting on a new set of fermions
which also carry Standard Model color charges. We expect this new
force to become confining at a scale $\Lambda_{HC}$, resulting in a
set of exotic mesons. Here one can consider an analogy with the
breakdown of chiral symmetry in the Standard Model. If we neglect
quark masses for the light quarks, there is a global $SU(3)_L \times
SU(3)_R \times U(1)_L \times U(1)_R$ symmetry among the up,
the down, and the strange quarks. Strong QCD interactions generate a
quark-antiquark vacuum condensate which spontaneously breaks this
symmetry down to $SU(3) \times U(1)$. In this simple picture
one would expect 9 Goldstone bosons from the 9 broken currents,
arranged in a flavor octet and a singlet. Realistically, the unequal
quark masses and electric charges make these symmetries only
approximate, especially when considering the relatively heavy
strange quark. Rather than massless particles, we have
pseudo-Goldstone bosons with the familiar pions as a light isospin
triplet and somewhat heavier kaons and etas filling out the nonet.
This division within the nonet is due primarily to the strange
quark. The isospin singlet is also heavier than the pions due to a
non-vanishing anomaly in the isosinglet-gluon-gluon diagram, meaning
this current is not conserved in QCD. These scalars also have
heavier, spin-1 counterparts in the $\rho$, $K^{*}$, $\omega$, and
$\phi$ mesons.

For the hypercolor model a similar analysis applies. Massless
hyperquarks would have a global left-right flavor symmetry
which is spontaneously broken by a hypercolor-driven condensate. In
the Standard Model case there is an approximate
remaining flavor symmetry due to the lightness of the up, down, and
strange quarks. In the present example, however, this
role is filled by the SM gauged $SU(3)$ color symmetry which remains
exact. Hence the left-right flavor breaking should
preserve a color nonet of pseudo-Goldstone bosons, assuming that any
 mass terms for the hyperquarks  are
suitably below the scale of symmetry breaking. This nonet can be
decomposed as a massive color octet and a singlet.
Similarly to the isospin singlet mentioned above, the color singlet
will have a non-vanishing anomaly with two hyper-gluons and
we do not expect it to be as light as the octet.
The lightest new states in the effective theory are thus a color
octet of scalars, which are designated ``hyper-pions''
($\tilde{\pi}$) in reference to their standard model
analogues. These states should have a set of spin-1 excitations
which fill out a massive color octet of vector bosons. These ``hyper-rhos''
($\tilde{\rho}$) are \Cblue{called} colorons in the naming convention we follow.

%
% Effective Lagrangian
%
Based on this model of chiral symmetry breaking, the authors of
Ref.~\cite{Kilic:2008pm} derive the following effective Lagrangian:
%
% Effective Lagrangian
%
\begin{eqnarray}
{\cal L}_{\rm eff}
& = & -\frac{1}{4}G_{\mu\nu}^a G^{a\mu\nu} +\bar{q}i\notD q
 -g_3\epsilon \trho^a_\mu\bar{q}\gamma^\mu T^a q  \nonumber\\
&  & +\frac{1}{2}(D_\mu \tpi)^a (D^\mu \tpi)^a -{M_{\tpi}}^2\tpi^a\tpi^a
 -\frac{1}{4}\trho_{\mu\nu}^a\trho^{a\mu\nu}
 +\frac{M_{\trho}^2}{2}\trho_\mu^a\trho^{a\mu} \nonumber\\
&  & -i g_{\trho\tpi\tpi}f^{abc}\trho^{a\mu}(\tpi^b D_\mu\tpi^c) -\frac{3{g_3}^2 \epsilon^{\mu\nu\rho\sigma}}{16\pi^2 f_{\tpi}}
\, {\rm Tr}[\tpi G_{\mu\nu}G_{\rho\sigma}] \nonumber\\
&  & +i\chi g_3\, {\rm Tr}[G_{\mu\nu}[\trho^\mu,\trho^\nu]]
   +\xi\frac{2i\alpha_3\sqrt{N_{\rm HC}}}{M_{\trho}^2}
\, {\rm Tr}[ {\trho^\mu}_\nu [G^\nu_\sigma,G^{\sigma\mu}] ] \, .
%
%{\cal L}_{\rm eff}&=\overline{q}i \notD q - \frac{1}{4}G_{\mu\nu}^a G^{a\mu\nu}-\
% \frac{1}{4}\tilde\rho_{\mu\nu}^a\tilde\rho^{a\mu\nu}
%+ \frac{m_{\tilde\rho}^2}{2}\tilde\rho_\mu^a\tilde\rho^{a\mu} -g_3*\varepsilon \
%\tilde\rho^a_\mu\overline{q}\gamma^\mu T^a q \\ &+
%\frac{1}{2}(D_\mu \tilde\pi)^a (D^\mu \tilde\pi)^a -m_{\tilde\pi}^2\tilde\pi^a\\
%tilde\pi^a  -i*\tilde g_{\rho\pi\pi}f^{abc}\tilde\rho
%^{a\mu}(\tilde\pi^b D_\mu\tilde\pi^c) \\ &- \frac{3g_3^2 \epsilon^{\mu\nu\rho\s\
%igma}}{16\pi^2 f_{\tilde\pi}}tr[\tilde\pi G_{\mu\nu}
%G_{\rho\sigma}]+i\chi g_3 tr[G_{\mu\nu}[\tilde\rho^\mu,\tilde\rho^\nu]] +
%\xi \frac{2i\alpha_3\sqrt{N}}{m_{\tilde\rho}^2}tr[\rho^\mu_\nu[G^\nu_\sigma,G^{\
%\sigma_\mu}]].
\end{eqnarray}
\Cred{For simplicity, we assume only one flavor of hyper-quark, Q so
that we have only one pseudo-Goldstone boson $\tilde{\pi}$ and one
$Q \bar{Q}$ bound state, $\tilde{\rho}$}. In the above equation
$G_{\mu\nu}$ and $q$ are Standard Model gluon and quark fields,
while $a$ is a color index. \Cred{$SU(N_{\rm HC})$ is the the
symmetry group of the hypercolor gauge interaction and we will take
$N_{HC}$ to be $3$} for simplicity. Based on this assumption, Kilic et
al. have derived most of the parameters in terms of a single
variable, $M_{\tilde{\rho}}$. 
% (Check references for more general case.
% \Cblue{WHAT REFERENCES?  I SUGGEST WE REMOVE THIS}) 
\Cblue{They find} 
$\varepsilon \simeq 0.2$, $g_{\tilde{\rho}\tilde{\pi}\tilde{\pi}}\simeq 6$, 
$M_{\tilde\pi} \simeq 0.3 M_{\tilde\rho}$, and
$f_{\tilde\pi} \simeq f_{\pi}\frac{M_{\tilde\rho}}{m_{\rho}}$
\Cblue{where $f_{\pi}\,=\,92$ MeV, the standard pion decay constant, 
and $m_{\rho}$ is the mass of the ordinary $\rho$ 
meson~\cite{Kilic:2008ub,Kilic:2009mi}}.

%\Cgreen{(For an estimate of these parameters in
%a more generalized case, see \cite{Kilic:2009mi}.)}

The first line of the Lagrangian contains Standard Model QCD, plus
the potential for quarks to couple to the $\tilde{\rho}$. \Cred{The
second line contains kinematic and mass terms for the
$\tilde{\pi}$ and  $\tilde{\rho}$ with $D_{\mu}$ representing the SM
covariant derivative. The third line contains the couplings of the
$\tilde{\rho}$ to two $\tilde{\pi}$'s and the coupling of
$\tilde{\pi}$ to two gluons} \Cblue{which provide} the primary decay routes of
interest to us \Cblue{because of} the large coupling strength $g_{\tilde{\rho}\tilde{\pi}\tilde{\pi}}$
and the relatively suppressed pion-gluon-gluon coupling. 
%These
%assumptions play a key role in the phenomenology of this model. 
The last two terms of equation (1) give additional gluon couplings 
to the $\tilde{\rho}$. They contain the parameters $\chi$ and $\xi$, 
which cannot be extrapolated from the Standard Model
analogy as they would vanish in the Abelian case.
% For the bulk of our paper we will choose $\chi =1$, \Cred{which is the the case in the topcolor
% model}
% , and $\xi =0$, a conservative choice.
\Cblue{Other models, such as the topcolor model require $\chi\,=\,1$ and $\xi\,=\,0$. Further these are
the only values for which any model is unitary.  After we compare the cross section $pp \to \tpi\tpi +X$ for a few values of $\xi$
and the cross section $pp \to \trho\trho +X$ for a few values of $\chi$ we will set $\xi$ and $\chi$ to the unitary values
for the rest of the paper (Sec. IV A and beyond).}

\section{General Phenomenology}

Based on the model outlined above, the hyper-pion couples to gluons 
via an anomalous term which we may think of as a triangular loop 
involving hyper-quarks. This is similar to the decay of 
the Standard Model $\pi^0$ into two photons. 
The hyper-pion decay width is given by
\begin{equation}
 \Cgreen{  \Gamma_{\tilde{\pi}\to gg} = \frac{15\alpha_s^2 M_{\tilde{\pi}}^3}{256 \pi^3 f_{\tilde{\pi}}^2}}.
\end{equation}

The $\tilde{\pi}$s will thus appear as a decay into two jets 
%at a collider 
and can potentially be produced singly via gluon fusion.
The $\tilde{\rho}$, on the other hand, couples to quarks through
coloron-gluon mixing, as well as to gluons for non-zero values of
$\xi$. However, its dominant decay mode is into two $\tilde{\pi}$s
due to the large coupling
$g_{\tilde\rho\tilde\pi\tilde\pi}$. Thus for a large range of
parameters the observable \Cblue{coloron} decay is four jets arising from
two hyper-pion resonances. For $\xi = 0$ the
$\tilde{\rho}$ decay width, as a function of
\Cred{$M_{\tilde{\rho}}$}, is

\begin{equation}
\Cred{   \Gamma_{\tilde{\rho}} \simeq 0.19  M_{\tilde{\rho}}},
\end{equation}
with a branching fraction to $\tilde{\pi}$s of \Cred{$B(\tilde{\rho
}\to \tilde{\pi}\tilde{\pi}) \simeq 95 \%$}. One can see that the
decay width for the $\tilde{\rho}$ relative to its mass is quite
broad, while that for the $\tilde{\pi}$ is narrow.

%------------------------
% FIG 1
%------------------------

\begin{figure}[htb]
\centering\leavevmode
\epsfxsize=3.2in
\epsfbox{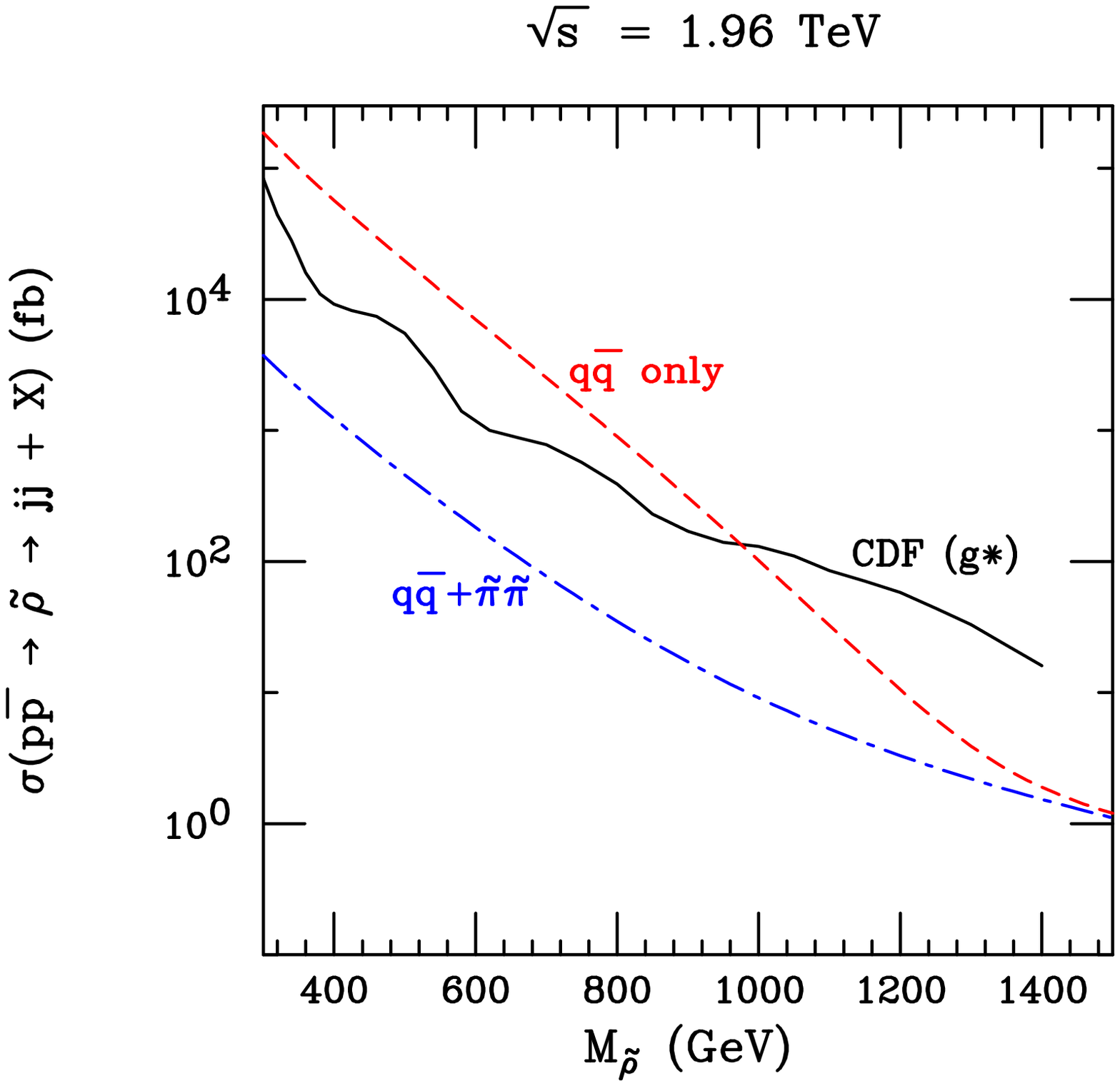}
\caption[]{
CDF exclusion limits (black solid line) for dijet-decaying color octets, 
compared to predicted $pp \to \trho \to jj+X$ cross sections computed 
with $\tpi$ decays (blue dot-dashed line) and 
without $\tpi$ decays (red dashed line) 
for $\sqrt{s} = 1.96$ TeV.
\label{fig:tevexcl}
}
\end{figure}

%------------------------
% FIG 2
%------------------------

\begin{figure}[htb]
\centering\leavevmode
\epsfxsize=3.2in
\epsfbox{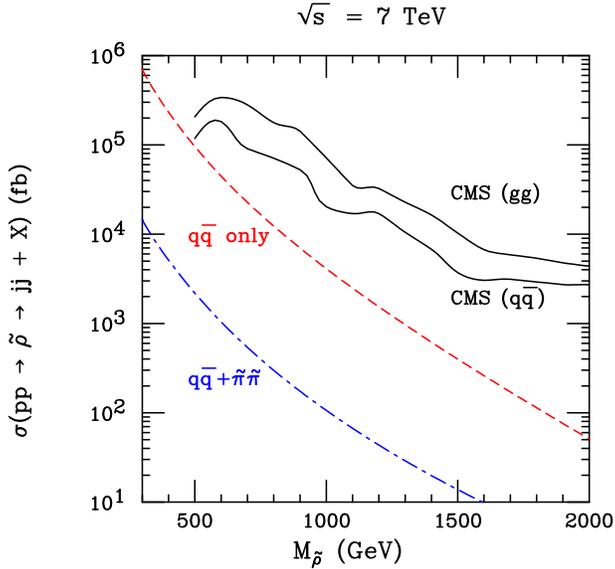}
\caption[]{
CMS exclusion limits (black solid line) for dijet-decaying color octets, 
compared to predicted $pp \to \trho \to jj+X$ 
cross sections computed 
with $\tpi$ decays (blue dot-dashed line) and 
without $\tpi$ decays (red dashed line) 
for $\sqrt{s} = 7$ TeV.
\label{fig:lhcexcl}
}
\end{figure}

The dijet decay modes of both the scalar and the vector octets provide 
possible constraints on the model because of the dijet resonance exclusion 
bounds from the Tevatron Run-II data~\cite{Aaltonen:2008dn,Zerwekh:2010uk}.
However, for the coloron, the branching ratio to dijets is small 
(less than $5$ percent), suppressing the signal below existing bounds. 
This would not be true without the strongly coupled pion decay mode, 
as can be seen in Fig.~\ref{fig:tevexcl}. In this figure the solid black line 
shows the bound for dijet decays of a coloron-like particle at the CDF
detector~\cite{Aaltonen:2008dn}.  
The dashed (red) curve represents the total cross section for 
$pp \to \trho \to jj +X$ in the absence of the pion decay mode.
The dot-dash (blue) line indicates the total dijet cross section with
this mode taken into account. 
One can see that without the large pion contributions to the $\trho$ width, 
colorons would be ruled out below about 1 TeV. 
As shown in Fig.~\ref{fig:lhcexcl} the LHC early searches 
for exotic dijet resonances~\cite{Khachatryan:2010jd}
are not yet competitive with the CDF bound~\cite{Aaltonen:2008dn}.
Hence, it will be difficult to observe a signal for the coloron 
via this dijet mode. 

% for the range of masses that we consider, dijet production
% cross section\Cgreen{s} at the Tevatron \Cgreen{are} sufficiently 
% small to remain
% currently undetected. In the case of a resonant hyper-pion this is
% because of the suppressed vertex required for production. 

%The current  bound for the
%$\tilde{\rho}$ mass from the CMS collaboration at the LHC via this
%dijet mode is xx GeV  \cite{xxxxxxx}}.
%$For the hyper-rho, the large branching
%fraction to two $\tilde{\pi}$s followed by four gluons is crucial. A
%large dijet branching fraction to two quarks (or two gluons) would
%likely be excluded for $\tilde{\rho}$ masses less than about  1 TeV.
%There are also
% early analyses of dijet searches at the LHC,
%but the exclusion bounds for exotic resonances are currently less stringent than those
% derived from Tevatron
%data.\cite{Khachatryan:2010jd}

\begin{figure}[htb]
\centering\leavevmode
\epsfxsize=3.4in
\epsfbox{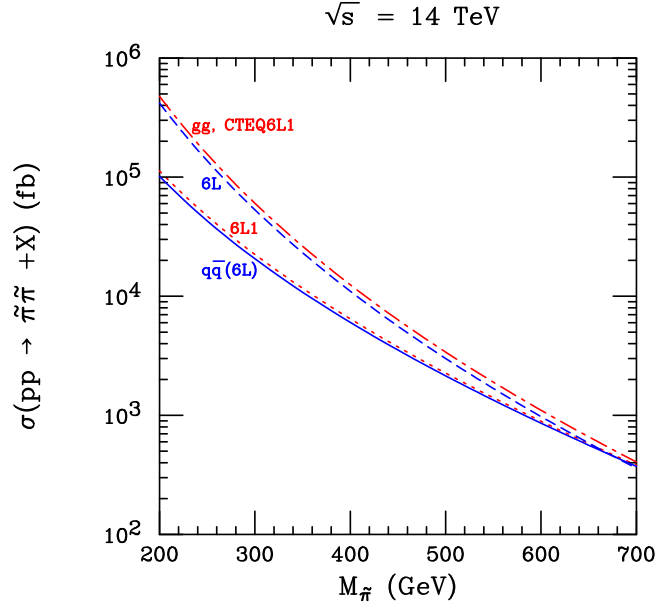}
\caption[]{
The cross section of $pp \to \tpi\tpi +X$ from $gg$ and $q\bar{q}$
%for $\sqrt{s} = 14$ TeV, 
as a function of $M_{\tpi}$.
We compare cross sections for two sets of parton distribution
functions, CTEQ6L1 and CTEQ6L.~\cite{CTEQ6}
\label{fig:ppto2pi1}
}
\end{figure}

\begin{figure}[htb]
\centering\leavevmode
\epsfxsize=3.4in
\epsfbox{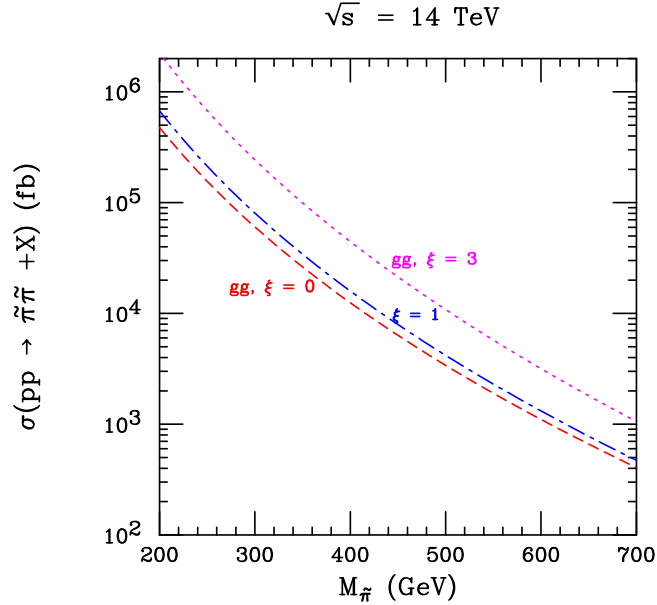}
\caption[]{
The cross section of $pp \to \tpi\tpi +X$ from gluon initial states,
as a function of $M_{\tpi}$,
for $\xi = 0, 1$, and $3$. 
%at the LHC with $\sqrt{s} = 14$ TeV.
%
\label{fig:ppto2pi2}
}
\end{figure}

A natural channel to search for the coloron is the resonant
$\tilde\rho$ production branching to two $\tilde\pi$'s which decay into
four jets. This mode has been studied \Cblue{for} the Tevatron
in Ref.~\cite{Kilic:2008pm} and \Cblue{for} the LHC
in Ref.~\cite{Kilic:2008ub}.
For comparison we present our results for the production cross section
$pp \to \tilde\pi \tilde\pi + X$ as a function of $M_{\tilde\pi}$
in Fig.~\ref{fig:ppto2pi1}.
In addition, Fig.~\ref{fig:ppto2pi2} shows the possible
enhancement of the signal with a non-zero value for $\xi$.
We note that our results appear to be \Cblue{smaller} than those shown in the
equivalent figure of Kilic et al. by approximately a factor of two.

%------------------------
% FIG 3
%------------------------

%------------------------
% FIG 4
%------------------------

\begin{figure}
\begin{center}
\begin{picture}(300,200)(0, 0)
\Gluon(5,180)(62,180){3}{4}
\Gluon(5,120)(62,120){3}{4}
\Gluon(62,180)(62,120){3}{4}
\DashLine(62,180)(120,180){3}
\DashLine(62,120)(120,120){3}
\Text(0,180)[t]{$g$}
\Text(0,120)[t]{$g$}
\Text(125,180)[t]{$\tilde \pi$}
\Text(125,120)[t]{$\tilde \pi$}
\Text(62,110)[t]{A}

\Gluon(155,80)(212,80){3}{4}
\Gluon(155,20)(212,20){3}{4}
\DashLine(212,80)(212,20){3}
\DashLine(212,80)(270,80){3}
\DashLine(212,20)(270,20){3}
\Text(150,80)[t]{$g$}
\Text(150,20)[t]{$g$}
\Text(275,80)[t]{$\tilde \pi$}
\Text(275,20)[t]{$\tilde \pi$}
\Text(221,10)[t]{D}

\Gluon(5,80)(62,50){3}{4}
\Gluon(5,20)(62,50){3}{4}
\DashLine(62,50)(120,80){3}
\DashLine(62,50)(120,20){3}
\Text(0,80)[t]{$g$}
\Text(0,20)[t]{$g$}
\Text(125,80)[t]{$\tilde \pi$}
\Text(125,20)[t]{$\tilde \pi$}
\Text(62,10)[t]{C}

\Gluon(155,180)(190,150){3}{4}
\Gluon(190,150)(155,120){3}{4}
\Gluon(190,150)(230,150){3}{4}
\DashLine(230,150)(270,180){3}
\DashLine(230,150)(270,120){3}
\Text(150,180)[t]{$g$}
\Text(150,120)[t]{$g$}
\Text(275,180)[t]{$\tilde \pi$}
\Text(275,120)[t]{$\tilde \pi$}
\Text(212,110)[t]{B}
\end{picture}
\end{center}
\caption[]{
Feynman diagrams of $gg \to \tpi\tpi$. 
\label{fig:feynman}
}
\end{figure}
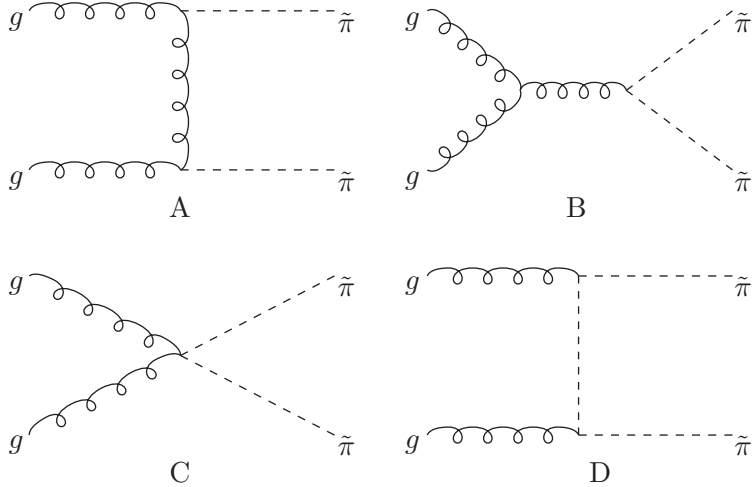

However, as pointed out in Ref.~\cite{Kilic:2008ub}, if our goal is
to distinguish the primary coloron resonance, this mode presents
some difficulties, particularly at the LHC. Although the $\tilde\pi$
resonances should be clear for accessible mass ranges, this may not
be the  case for the \Cblue{coloron}. While the hyper-pion decay width is
quite small compared to its mass, leading to a sharp peak in the
pair invariant $2$-jet mass distribution, the $\tilde\rho$ has a
large width ($\sim 0.2 M_{\tilde\rho}$) leading to a broad peak.
Moreover, there are important alternate modes of hyper-pion pair
production in this model, 
which do not involve the coloron, as shown in Fig.~\ref{fig:feynman}.

%------------------------
% FIG 5
%------------------------

The first graph of Fig.~\ref{fig:feynman}, A, is negligible
owing to the smallness of the $g-g-\tilde\pi$ vertex
but the others comprise a significant part of the two-$\tilde\pi$ signal.
These contributions derive from two initial gluons,
thus they are especially important at the LHC.

\begin{figure}[htb]
\centering\leavevmode
\epsfxsize=3.4in
\epsfbox{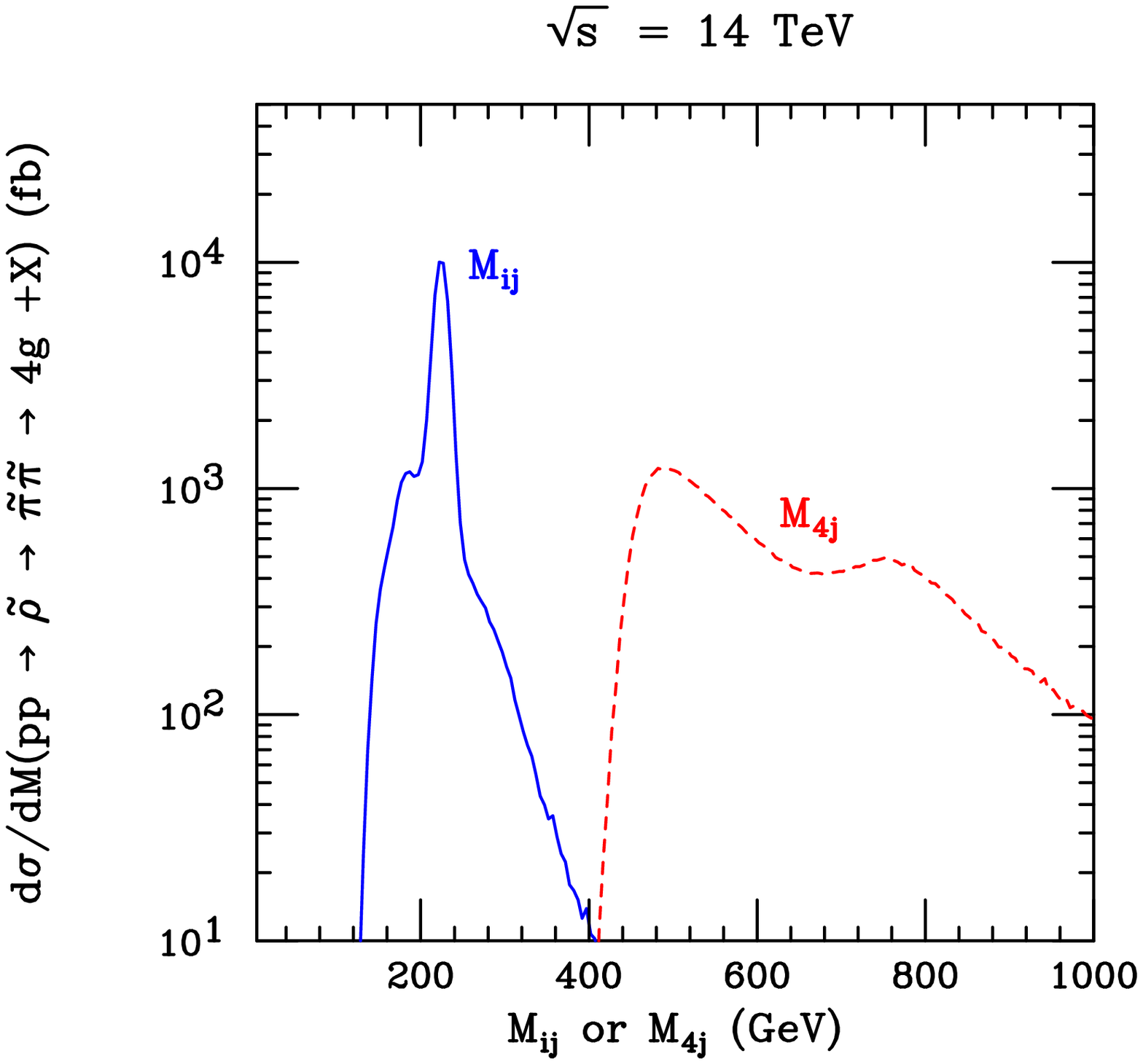}
\caption[]{
Invariant mass distribution of two jets ($M_{jj}$) or four jets ($M_{4j}$)
from $pp \to \trho \to \tpi\tpi \to 4g +X$. 
%with $\sqrt{s} = 14$ TeV.
%
\label{fig:bto4gmi}
}
\end{figure}

The importance of non-resonant diagrams can be seen in Fig.~\ref{fig:bto4gmi}.
This plot was produced for $M_{\trho} = 750$ GeV
($M_{\tpi} = 225$ GeV) for $\sqrt{s} = 14$ TeV.
Only minimal cuts were applied; all jets were required to have
$p_T > 15$ GeV and $|\eta| < 2.5$ and all pairs of jets were required to be
separated by $\Delta R > 0.5$.
We consider the signal $pp \to \trho \to \tpi\tpi \to 4g +X$
and apply a Gaussian smearing routine to outgoing momenta.
There are 3 ways of pairing up the 4 final jets and we require that
at least one \Cblue{such pairing} results in two pair masses within 50 GeV of each other.
The solid (blue) curve represents the invariant mass of a pair of jets.
This curve shows the average mass obtained thereby and one can
clearly see the hyper-pion peak.
(If more than one arrangement allows two nearby pair masses, we
average over these as well.) The signal reconstruction away from the
main peak can be reduced by narrowing the 50 GeV window.

%------------------------
% FIG 6
%------------------------

%------------------------
% FIG 7
%------------------------

\begin{figure}[htb]
\centering\leavevmode
\epsfxsize=3.4in
\epsfbox{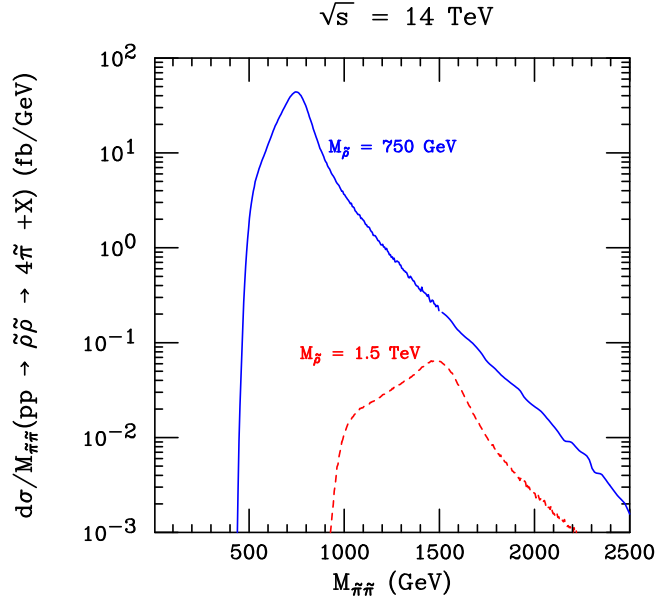}
\caption[]{
Invariant mass distribution of two $\tpi$'s
for $pp  \to 4\tpi +X$ with $\sqrt{s} = 14$ TeV. 

\label{fig:4pipeak}
}
\end{figure}

The dashed (red) curve shows the $4-$jet invariant mass constructed
from the sum of the final state momenta. 
The peak is not located at the $750$ GeV region; in fact the
$\tilde\rho$ mass appears as only a broad local maximum. 
It is difficult to improve the coloron detection in this
channel without foreknowledge of the $\tilde\pi$-$\tilde\rho$ mass relation.

For this reason we focus in this paper on an alternate detection channel.
We investigate the possibility of pair-produced colorons at the LHC,
$pp \to \trho\trho \to 4\tpi \to 8g +X$.
Pair production of $\tilde\rho$'s receives contributions from both $gg$
and $q\bar{q}$ initial states.
More importantly, the presence of two identical massive resonances allows
us a better chance of distinguishing them by correlating 4-jet
invariant masses. Fig.~\ref{fig:4pipeak}  was calculated from the
four hyper-pion signal at the LHC. Similarly to the hyper-pion signal
represented in Fig.~\ref{fig:bto4gmi},
we simulate events and reconstruct an average mass for two pairs
of hyper-pions with invariant masses within 100 GeV of each other.
The peak, shown with a solid (blue) or a dashed (red) curve, is broad 
but it is clearly located at the chosen $\tilde\rho$ mass of 750 GeV or 
1500 GeV, \Cblue{even though we have not made any}
assumptions \Cblue{to influence that value} in our reconstruction. 
Despite the fact that the calculated signal, $pp \to 4 \tilde\pi + X$ 
includes hundreds of graphs without any intermediate $\tilde\rho$s, 
only those which \Cblue{do have intermediate $\tilde\rho$s} are
likely to contribute significantly to the correlated signal. The peak
at the correct mass should improve with a narrower window, but we must
be careful not to restrict it too much. The inherently broad width of
the $\tilde\rho$, compounded by the resolution limitations of a real
detector caution against killing our signal with over-fine mass searches.

%\Cred{Comment: State in the figure 5 caption  what is the red curve for}. 
%\Cblue{Done}

\section{Signal Simulation}

Our chosen signal is $pp \to \trho\trho \to 4\tilde\pi \to 8g + X$. 
In order to run Monte-Carlo simulations of events we make use of 
MadGraph II~\cite{Stelzer:1994ta,Alwall:2007st} to generate the
squared matrix elements. 
We have built the coloron model into the MadGraph/MadEvent4 code 
using the available ``user model'' framework. 
However, to implement the full model we have made some modifications 
to the underlying code for generating color factors and we have added 
some additional routines to the HELAS library~\cite{Murayama:1992gi} 
used by MadGraph. This is necessitated by the non-Standard-Model vertices 
which facilitate some of our interactions. 
The $g-\tilde\rho-\tilde\rho$ vertex, for example, is similar to the 
Standard Model 3-gluon interaction, but contains a $\chi$ dependence 
for the terms involving the gluon momentum.

%\Cblue{NEED TO PUT THIS PARAGRAPH SOMEWHERE, HOW ABOUT HERE}

The authors of refs \cite{Kilic:2008pm} and \cite{Kilic:2008ub}
implemented the model in the AMEGIC matrix element generator which
is used by the SHERPA event generator. We have chosen to build the
model into MadGraph so as to have an independent calculation of the
relevant cross sections. We calculate our backgrounds with SHERPA.
%For 14 TeV c.m. energies we find good detection prospect for
%hyper-pion/hyper-rho masses out to 450 GeV/1.5 TeV \Cred{with $30
%fb^{-1}$ of luminosity}.

In the case of the $pp \to 8g +X$ decay chain, 
we had to make some modifications to handle the very large terms which 
arise in the color factors. 
The complete $pp \to 8g +X$ matrix element with all permutations of 
virtual particles is beyond the powers of MadGraph4 to generate 
due to the large number of graphs. Fortunately, the $\tilde\pi$ has 
a very small width relative to its mass, making it an excellent candidate 
for the Narrow Width Approximation (NWA). 
This small width means that interference terms between different 
arrangements of virtual hyper-pions are small compared to the resonant 
contributions. Thus we are able to generate the matrix element
$pp \to (\trho \to (\tpi \to gg)(\tpi \to gg))
        (\trho \to (\tpi \to gg)(\tpi \to gg)) +X$ 
including the $\tilde\pi$ width.
This  should be an excellent approximation to the complete calculation, 
and checking the $pp \to 2\tilde\rho \to 4\tilde\pi +X$ matrix element 
with on-shell hyper-pion decays we do find very good agreement. 
Thus the NWA for the $\tpi$s is valid and we make use of it for 
our signal generation. The width of the $\tilde\rho$ on the other hand 
makes the NWA unsuitable for that particle and we keep its width 
and interference terms throughout.

To further check our MadGraph results we have done an independent analytic calculation of the $gg \to 2\tilde\rho \to 4\tilde\pi$ matrix element,
including the $\chi$ and width dependence. Tested with realistic momenta, we have near-perfect concordance of numerical results from the two
computations.

For completeness we present here our analytic formula for the square of 
the matrix element $gg \to \tilde\rho \tilde\rho$, 
summed over polarizations and colors, as a function of  $\chi$. 
We neglect the coloron width as it would make for a substantially 
longer expression.
\begin{eqnarray}
\sum_{{\rm pol}}|T|^2\,
 &=&\,\frac{Y^2(1-z^2)^2}{(1-\beta^2z^2)^2}\frac{E^4}{M_{\trho}^4}
     \left[12-12Y+(5+z^2)Y^2\right]  \nonumber \\
 &+&\,\frac{Y^2(1-z^2)}{(1-\beta^2z^2)^2}\frac{E^2}{M_{\trho}^2}
     \left[16(1+3z^2)-2(11+18z^2)Y+(5+9z^2+3z^4)Y^2\right]  \nonumber \\
 &+&\,\frac{1}{(1-\beta^2z^2)^2}\bigg[8\left(16+3\frac{M_{\trho}^4}{E^4}\right)
      -256Y+(160+16z^2+36z^4)Y^2  \nonumber  \\
 &&\,\,\,\,\,\,\,\,\,\,\,\,\,\,\,\,\,\,\,\,\,\,\,\,\,\,\,\,\,\,\,\,\,\,\,\,\,\,\,\,\,\,\,\,\,\
\,-(32+22z^2+24z^4)Y^3+(2+5z^2+4z^4+2z^6)Y^4\bigg]  \nonumber  \\
 &+&\frac{1}{1-\beta^2z^2}\bigg[-6\left(16+4\frac{M_{\trho}^2}{E^2}
    +\frac{M_{\trho}^4}{E^4}\right)+140Y
    -(58+24z^2)Y^2 \nonumber \\
 &&\,\,\,\,\,\,\,\,\,\,\,\,\,\,\,\,\,\,\,\,\,\,\,\,\,\,\,\,\,\,\,\,\,\,\,\,\,\,\,\,\,\,\,\,\,\,\,\,\,\,\,\,\,\,\,\,\,\,\,\,
\,\,\,\,\,\,\,\,\,\,\,\,\,\,\,\,\,\,\,\,\,\,\,\,\,\,\,\,\,\,\,\,\,\,\,\,\,\,\,\,\,\,\,\,\,\,\,\,\,\,+3(1+4z^2)Y^3-z^4Y^4\bigg]  \nonumber \\
 &+&\,28+6\frac{M_{\trho}^2}{E^2}-3(1-\beta^2z^2)-16Y+4Y^2\,,
\end{eqnarray}
where $E$ is the gluon c.m. energy,
$z$ is the cosine of the scattering angle,
$\beta^2\,=\,1-M_{\trho}^2/E^2$,
and $Y\,=\,1-\chi$.

It is interesting to compare this to another calculation of coloron
pair production in a somewhat different model, given in
Ref.~\cite{Dicus:1994sw}. In that model, colorons arise from the
spontaneous breaking of an \Cred{$SU(3)_I\times SU(3)_{II}$ gauge
symmetry to the familiar $SU(3)_{color}$.} \Cblue{(}  \Cred{In the topcolor model,
the $SU(3)_I$ couples to the first two families  of the SM
fermions, while the $SU(3)_{II}$ couples to the 3rd family}.\Cblue{)} This
results in a set of massive partners to the SM gluons through a
Higgs mechanism, and these colorons are similar to the ones we
discuss. However, they do not have the $\chi$ dependence shown above
and we have checked that our formula is equivalent for the case
$\chi = 1$. \Cred{Recall that in the model we consider, the colorons
are not the product of a broken gauge symmetry, but are composite
particles ($Q\bar{Q}$) of the hyper-quarks}.
%with relatively small masses due to a broken
%approximate global symmetry.
 In the limit where $\chi \to 1$ the
$\tilde\rho$s couple to gluons in exactly the same fashion as other
gluons, so perhaps it is not too surprising that they then mimic the
``hyper-gluons'' of the model of Ref.~\cite{Dicus:1994sw}.

Only for $\chi = 1$ is the theory given in Eq.~(1) explicitly unitary 
for $gg \to \tilde\rho\tilde\rho$. This can be seen in Eq.~(4) above 
where the first lines grow \Cblue{with energy} unless $Y=0$.
In general
the coloron model we consider is only an effective theory which results from integrating out heavy,
 strongly-interacting hyper-quarks. The Lagrangian
includes non-renormalizable terms and does not include any explicit hyper-color gauge fields, nor does it
explain the origin of hyper-quark
masses. Thus as a theory of massive vector bosons we do not expect
it to preserve unitarity to all energies. In the $\chi = 1$ case,
however, the $\tilde\rho$s have gluon couplings equivalent to the
colorons of spontaneously broken $SU(3)_I \times SU(3)_{II}$. In
that model we think of the longitudinal components of the massive
bosons as coming from  `eaten' Higgs fields, and in the high-energy
limit unitarity is restored by the Goldstone Boson Equivalence theorem 
and the underlying gauge invariance, 
just as in the case of $W^+ W^-$ scattering.

\begin{figure}[htb]
\centering\leavevmode
\epsfxsize=3.4in
\epsfbox{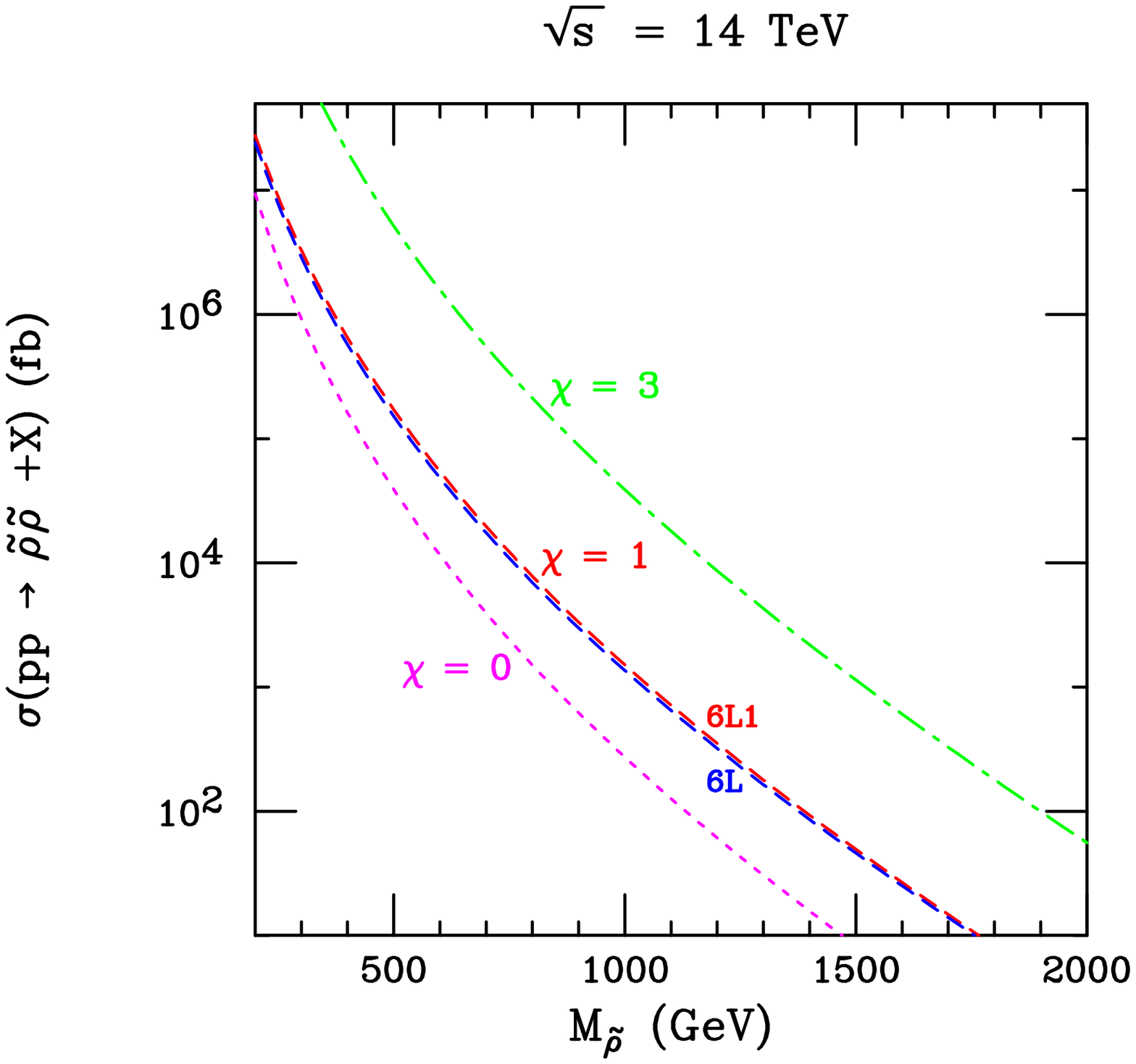}
\caption[]{
The cross section of $pp \to \trho\trho +X$, as a function of $M_{\trho}$, for $\chi = 1, 0$, and $3$.
%at the LHC with $\sqrt{s} = 14$ TeV.
%
\label{fig:ppto2b}
}
\end{figure}

In Fig.~\ref{fig:ppto2b} we show the total two-$\tilde\rho$
production cross section as a function of the coloron mass. We
include the prediction for $\chi = 0,\,1,\,\text{and}\,3$ to demonstrate
the effect of this parameter. 
\Cgreen{This figure may be compared to the results 
shown in Ref.~\cite{Kilic:2008ub}. Our estimates are broadly consistent. 
However, for some parameters their predicted cross section appears 
to be as much as twice our result. The discrepancy is more pronounced
at high values of $M_{\trho}$ and for non-unitary values of $\chi$.} 
Henceforth we will use $\chi$ equal to one.

%(Do we want to explicitly compare with Hopkins here? Effects of $\xi$ ?)
%\Cred{ Comment: Yes, compare with the Hopkins group's result. Also
%discuss the effect of $\xi$ }

\begin{figure}[htb]
\centering\leavevmode
\epsfxsize=3.2in
\epsfbox{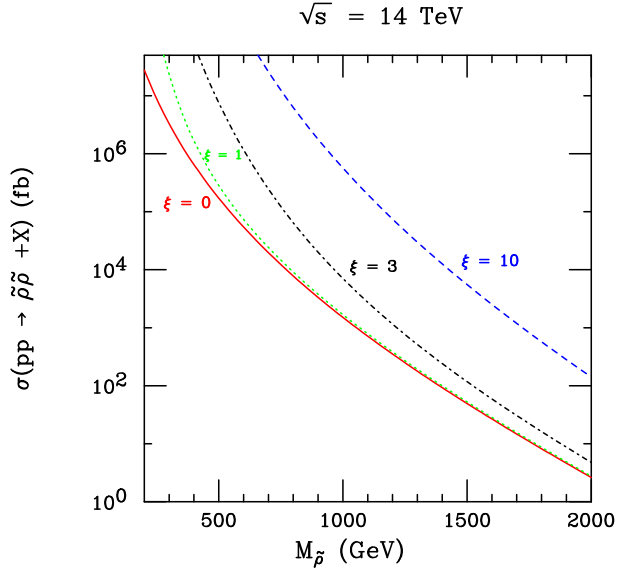}
\caption[]{
The cross section of $pp \to \trho\trho +X$, as a function of $M_{\trho}$, for $\xi = 0, 1, 3$, and $10$
at the LHC with $\sqrt{s} = 14$ TeV.
\label{fig:xiplot}
}
\end{figure}

\Cgreen{
The parameter $\xi$, which scales the $g-g-\trho$ vertex, can enhance 
the signal if it has a non-zero value. This vertex derives from 
a non-renormalizable term in the effective Lagrangian, 
which is the lowest order term allowing a direct gluon-$\trho$ coupling. 
Its effect on the two $\trho$ signal is shown in Fig.~\ref{fig:xiplot} 
for $\xi =1$ (dotted, green), $\xi=3$ (dot-dash, black) and
$\xi=10$ (dashed, blue) compared to $\xi =0$ (solid, red). 
Order one values of $\xi$ would significantly enhance the signal 
in the lower mass range. The effect is insensitive to the sign of $\xi$. 
(The four-jet, single-$\trho$ channel would also be enhanced, with 
potentially better resolution of the $\trho$ resonance since $\xi$ 
does not contribute to the non-resonant pion-pair graphs.)}

\begin{figure}[htb]
\centering\leavevmode
\epsfxsize=3.2in
\epsfbox{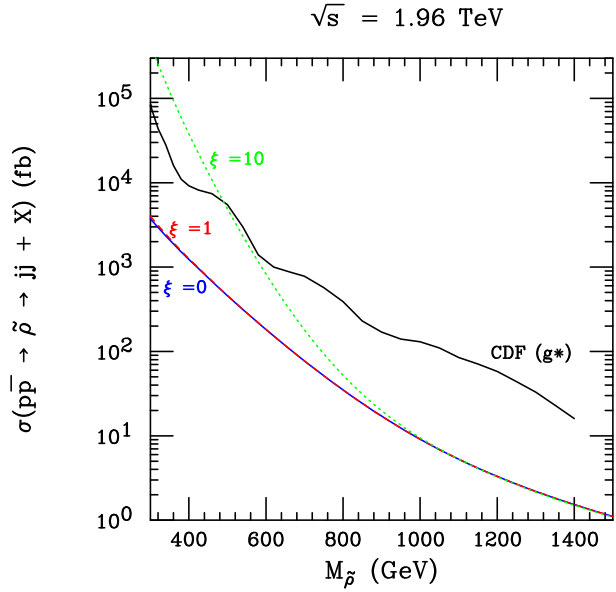}
\caption[]{
The cross section of $pp \to \trho \to jj +X$ for $\xi = 0, 1$, and $10$, 
compared to exclusion bounds at the Tevatron CDF detector.
\label{fig:exclxi}
}
\end{figure}

\Cgreen{
On the other hand, increasing $\xi$ also affects the dijet signal and 
potentially runs afoul of the Tevatron exclusion bound, as shown 
in Fig.~\ref{fig:exclxi}. For $\xi=1$ (dash, red) the dijet 
prediction is virtually the same as for $\xi=0$ (solid, blue). 
However if $\xi$ is allowed to be as large as $10$ (dotted, green), 
then the current bounds require $M_{\trho}$ to be greater than 
a few-hundred GeV. For the remainder of the paper we will set $\xi=0$, 
the most conservative choice.
}

\subsection{Signal Selection}

We proceed to generate signal samples using our MadGraph-generated
matrix elements and decaying the hyper-pions via the NWA. These are
convoluted with the CTEQ6L1 parton distribution functions. For the
signal we have followed the prescription that we use the mass of the
pair produced particle as the factorization and renormalization
scale, i.e. $\mu_F = \mu_R = M_{\tilde\rho}$. We take the K-factor
to be one. We apply a Gaussian smearing routine to the outgoing
momenta based on ATLAS specifications~\cite{ATLAS}

\begin{eqnarray}
\frac{\Delta E}{E} = \frac{0.60}{\sqrt{E({\rm GeV})}} \oplus 0.03\,\,.
\end{eqnarray}

\begin{figure}[htb]
\centering\leavevmode
\epsfxsize=4.2in
\epsfbox{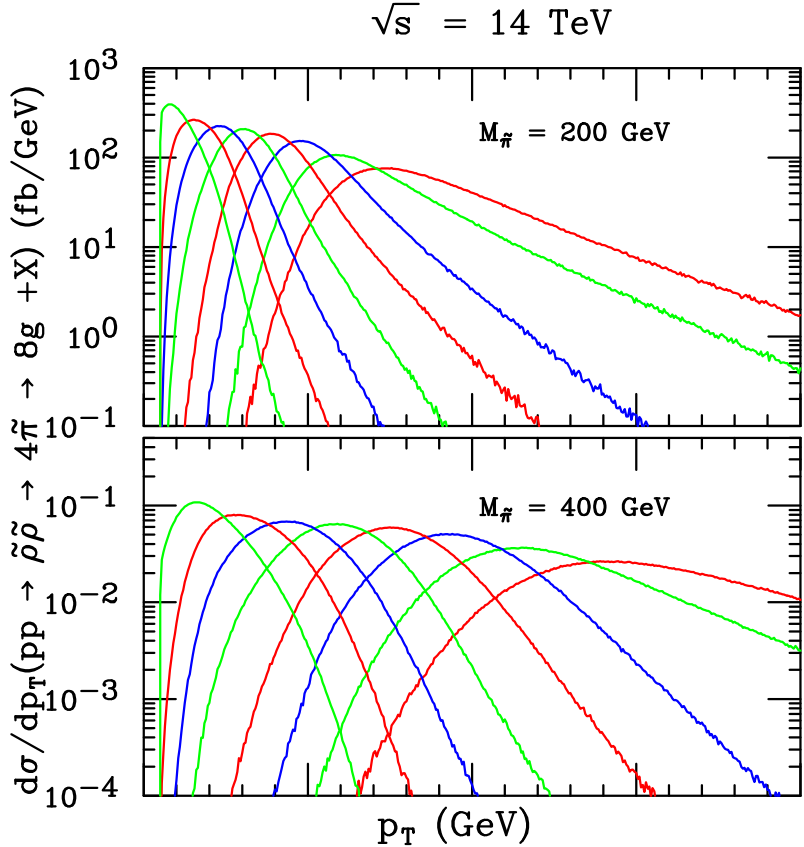}
\caption[]{
Transverse momentum ($p_T$) distributions of the eight signal gluons for 
$M_{\tpi} = 200$ GeV (top) and $M_{\tpi} = 400$ GeV (bottom).
\label{fig:clrpt}
}
\end{figure}

In all cases we require that the final gluons are within $-2.5 < \eta < 2.5$ and separated from one another by $\Delta R > 0.5$, so we assume they are
reconstructed as eight jets. These jets should have a high $p_T$ distribution in general due to boosts from the heavy $\tilde\rho$ and $\tilde\pi$
decays. Ordered by $p_T$, the outgoing gluon $p_T$ profiles present a succession of peaks, \Cgreen{as shown in Fig.~\ref{fig:clrpt}. We find that} the leading $p_T$ gluon typically peaks around
$\frac{M_{\tilde\rho}}{2} \simeq 1.5 M_{\tilde\pi}$.
%\Cred{ Do we have a figure here the substantiate the statements in
% the above two sentences?}
Particularly for high coloron masses, high $p_T$ cuts on the leading
jets provide strong discrimination against the background.

Aside from transverse momentum requirements, our primary tool for reducing background is selection for invariant masses. We have pursued two different
cut algorithms for this purpose.

\subsection{Relative Mass Windows}

For {\it relative} mass cuts we require that there is an arrangement of 
the 8 gluons into 4 pairs such that the largest and smallest invariant 
masses of these pairs are within a given window of one another 
($\Delta M_{ij}$). 
We then require that these candidate hyper-pions can be arranged 
into pairs such that the two pion-pair (4-gluon) invariant masses are 
within a window $\Delta M_{4j}$.
A successful arrangement of the gluons into candidate hyper-pions 
and \Cblue{colorons} must satisfy both mass window \Cblue{requirements} 
simultaneously. For example, a set of gluons $1,\ldots,8$ which passes 
the pion cut with the arrangement $(12)(34)(56)(78)$ can pass the rho 
cut with 4-masses such as $(1234)(5678)$ but not with $(1235)(4678)$.

\begin{figure}[htb]
\centering\leavevmode
\epsfxsize=3.4in
\epsfbox{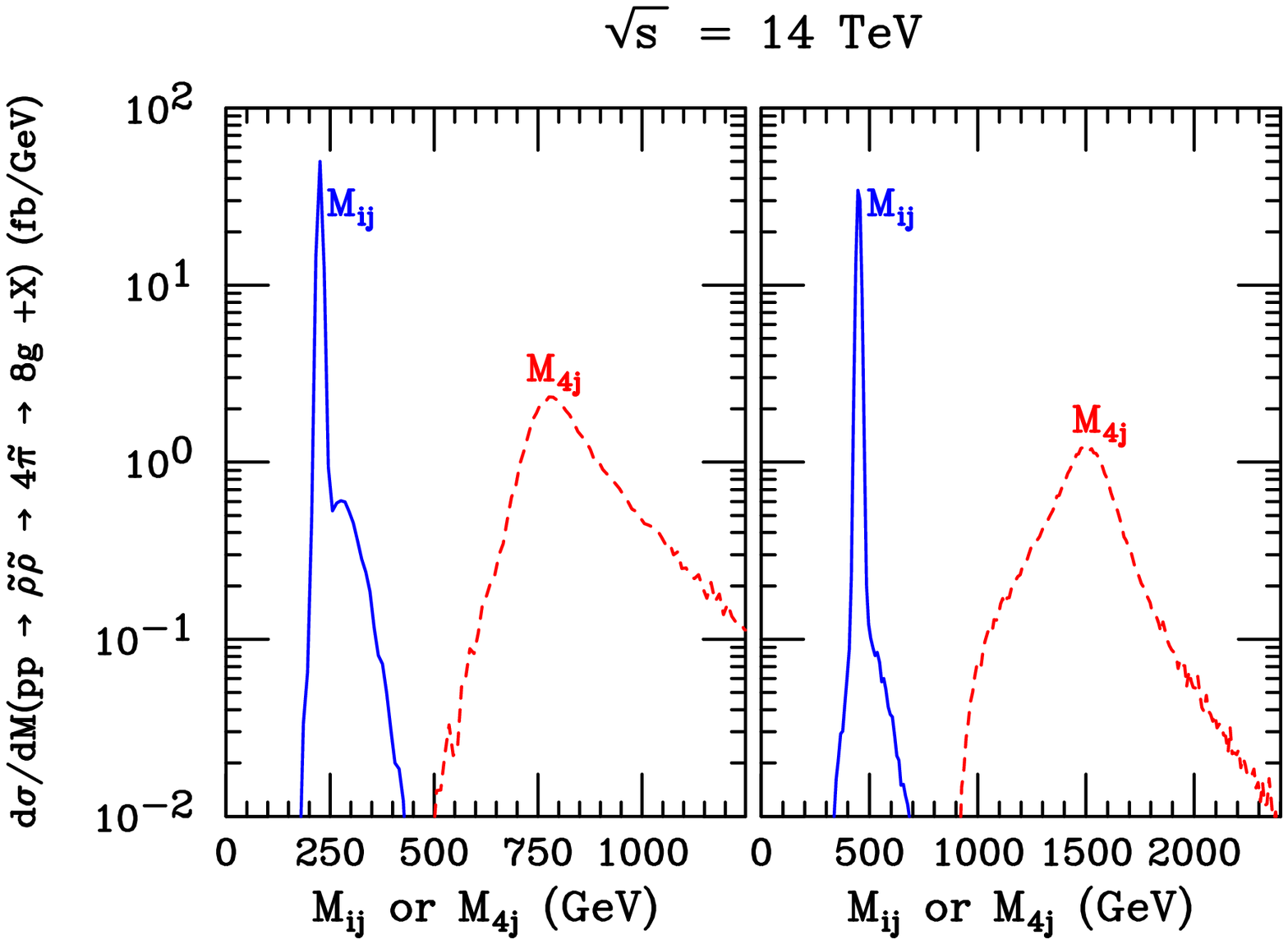}
\caption[]{
Invariant mass distribution of two jets or four jets
for $pp \to \trho\trho \to 4\tpi \to 8g +X$.
% with $\sqrt{s} = 14$ TeV.
%
\label{fig:8gpeaks}
}
\end{figure}

Fig.~\ref{fig:8gpeaks} shows a plot of the signal for $M_{\tpi} = 225$ GeV 
and $M_{\trho}=750$ GeV based on relative window mass cuts with 
$\Delta M_{ij} = 50$ GeV and $\Delta M_{4j} = 100$ GeV. 
Minimal momentum cuts of $p_T > 15$ GeV are applied. 
The solid (blue) line shows the average of candidate hyper-pion masses 
which pass the cuts, and it is sharply peaked at the hyper-pion mass. 
The dashed (red) line shows the average candidate \Cblue{coloron} mass 
which is peaked at the \Cblue{coloron} mass as expected. 
The equivalent lines corresponding to physical masses $M_{\tpi} = 450$ GeV 
and $M_{\tilde\rho}=1.5$ TeV are shown with solid (blue) and dashed (red) 
curves respectively.

\subsection{Fixed Mass Windows}

For \Cblue{{\it fixed} mass cuts} we require a suitable arrangement of 
gluons into candidate $\tilde\pi$s and $\tilde\rho$s as above. But rather than
requiring that the reconstructed masses are \Cblue{each} within some window of
one another, we require that they all fall within a window around a
set mass. That is, we require all four $M_{ij}$s to satisfy
$|M_{ij}-M_{\tilde\pi}| < \Delta M_{ij}$ and the 4-gluon invariant masses 
corresponding to pairs of candidate hyper-pions to satisfy
$|M_{4j} - M_{\tilde\rho}| < \Delta M_{4j}$ for a chosen value of 
$M_{\tilde{\pi}}$ or $M_{\tilde{\rho}}$.

The advantage of the relative mass scheme is that it requires no prior assumptions about the $\tilde\pi$ or $\tilde\rho$ masses. It is based only on
correlations between the invariant masses within an experimental data set. On the other hand it is effectively sampling across all possible masses so
it is not as efficient as we might like for eliminating backgrounds. The fixed mass scheme will perform better against background if the parameters
$M_{\tilde\pi}$ and $M_{\tilde\rho}$ are chosen close to the actual physical masses of the particles. Probable values for these parameters can, for
example, be read off a plot similar to Fig. \ref{fig:8gpeaks} derived from a relative mass window analysis. Or we may imagine that the hyper-pion mass is established
by sliding a fixed pair-invariant-mass window to find signal over background excesses in the 8-jet or 4-jet channels. One could then test choices of
$M_{\tilde\rho}$ for a fixed value of $M_{\tilde\pi}$. 

%\Cblue{NOT SURE WHAT THIS NEXT SENTENCE MEANS}
% Our intent is to demonstrate that the signal and its characteristic 
% resonances is potentially observable with the general relative mass cuts, 
% and that the background can be greatly reduced with a more fine-grained 
% search and judicious choices of $p_T$ cuts.

\section{Background Simulation}

The background for our signal is 8 jets coming from Standard Model QCD 
processes. Obviously it is quite large and complex before $p_T$ and invariant
mass cuts. MadGraph/MadEvent4 cannot simulate more than 5 outgoing gluons with all terms included. Instead we rely on the SHERPA 1.2.2
~\cite{Gleisberg:2003xi} event
generator. SHERPA makes use of the COMIX~\cite{Gleisberg:2008fv} matrix element calculator, 
which applies color-dressed Berends-Giele recursion relations~\cite{Berends:1987me}. For numerical
calculations of high-multiplicity tree-level diagrams this seems to be the most efficient method currently available~\cite{Duhr:2006iq}. 
We again use CTEQ6L1 PDFs but with
 the renormalization and factorization scales set to $\sqrt{\langle p_T^2 \rangle}$, the root mean transverse momentum squared. We take the K factor to be one. Cuts are
applied as described for the signal.

\begin{figure}[htb]
\centering\leavevmode
\epsfxsize=3.6in
\epsfbox{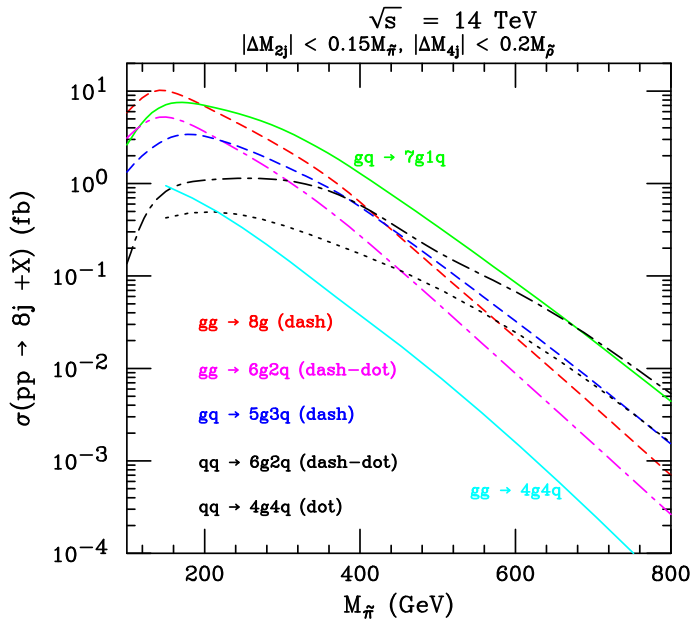}
\caption[]{
Individual background channels for $pp \to 8j + X$, calculated with
fixed mass window cuts and 
$p_T(j_1) > 1.5M_{\tpi}, \, p_T(j_2) > 1.2M_{\tpi}, \, p_T(j_3) > M_{\tpi}, 
\, p_T(j_4) > 0.8M_{\tpi}, \, p_T(j_5,j_6,j_7,j_8) > 50 \, {\rm GeV}$.
\label{fig:subbkgrd}
}
\end{figure}

The dominant backgrounds we consider are QCD processes for 
$pp \to 8j +X$ from $gg \to 8g$, $gq \to 7g1q$, $gg,qq \to 6g2q$, 
and $gq \to 5g3q$, where $q$ may be a quark or an antiquark. 
We expect these to account for the bulk of the background since
diagrams with higher numbers of quarks have relatively suppressed color
factors and fewer graphs. Of the backgrounds we compute, which process 
dominates depends on the cuts we choose and the mass of the hyper-mesons. 
\Cgreen{This behaviour can be seen in Fig. \ref{fig:subbkgrd}.} 

At low masses where we generally consider lower $p_T$ cuts,  
the $gg \to 8g$ (dash, red) channel tends to be largest, followed by 
$gq \to 7g1q$ (solid, green) and $gg \to 6g2q$ (dash-dot, magenta).
As we move to higher masses and $p_T$ requirements, valence quarks 
in the initial state become more important since they are favored when 
the kinematics require a large fraction of the incoming proton's momentum. 
Thus $gq \to 7g1q$ becomes the largest contribution for much of our range 
while $gg \to 8g$, $gq \to 5g3q$ (dash, blue), and 
$qq \to 6g2q$ (dash-dot, black) are subdominant with comparable values 
and $gg \to 6g2q$ becomes an increasingly small fraction of the total. 
For very high masses and cuts, $qq \to 6g2q$ becomes the largest 
background since it is the only one which can have two up or 
two down quarks in the initial state. 
We have found that the background $qq \to 8g$ is consistently smaller 
than the dominant processes by more than an order of magnitude and 
we neglect it in our results. 

\Cgreen{We also include the next two possibly significant backgrounds 
in the figure, $qq \to 4g4q$ (dot, black) and $gg \to 4g4q$ (solid, cyan).} 
They remain well below the dominant channels for observable $M_{\tpi}$ 
and we do not include them in the background estimates.

%------------------------------------------------
% Discovery Potential at the LHC
%------------------------------------------------
\section{Discovery Potential at the LHC}

\begin{figure}[htb]
\centering\leavevmode
\epsfxsize=3.2in
\epsfbox{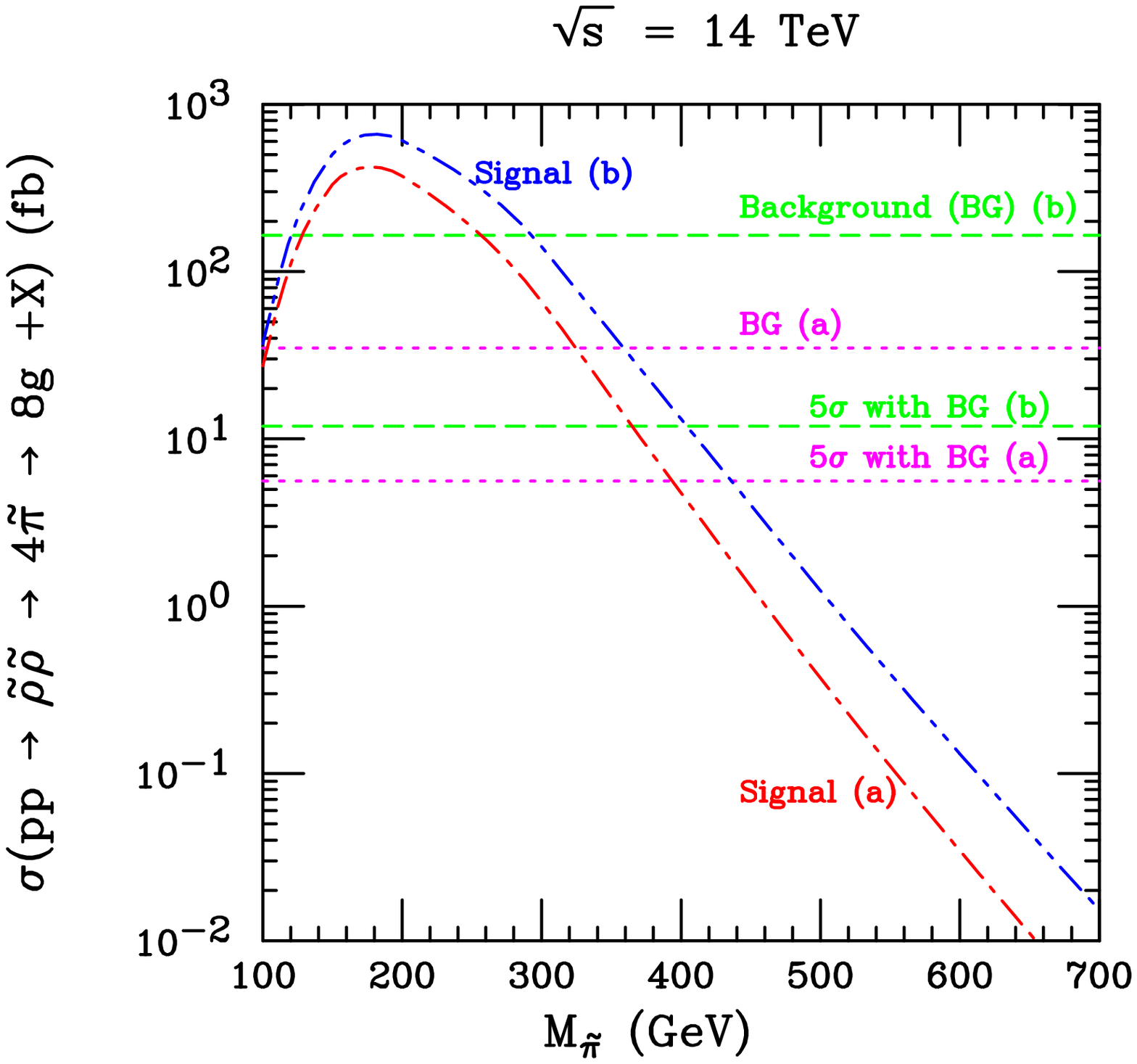}
\caption[]{
The cross section for $pp \to \trho\trho \to 4\tpi \to 8g +X$
%at the LHC with $\sqrt{s} = 14$ TeV, 
as a function of $M_{\tpi}$.
We have applied all kinematic cuts and two sets of
relative mass cuts:
(a) $\Del M_{2j} < 30$ GeV and $\Del M_{4j} < 60$ GeV 
    [red dot-dashed line], or
(b) $\Del M_{2j} < 50$ GeV and $\Del M_{4j} < 100$ GeV 
    [blue dot-dot-dashed line].
Also shown are the cross section for the dominant SM background 
with relative mass cut (a) [magenta dotted line] or (b) [green dashed line] 
as well as the minimal signal cross section that is required by
a 5 sigma criterion with relative mass cut (a) [magenta (lower) dotted line] 
or (b) [green (lower) dashed line].
\label{fig:relative}
}
\end{figure}

\begin{figure}[htb]
\centering\leavevmode
\epsfxsize=3.2in
\epsfbox{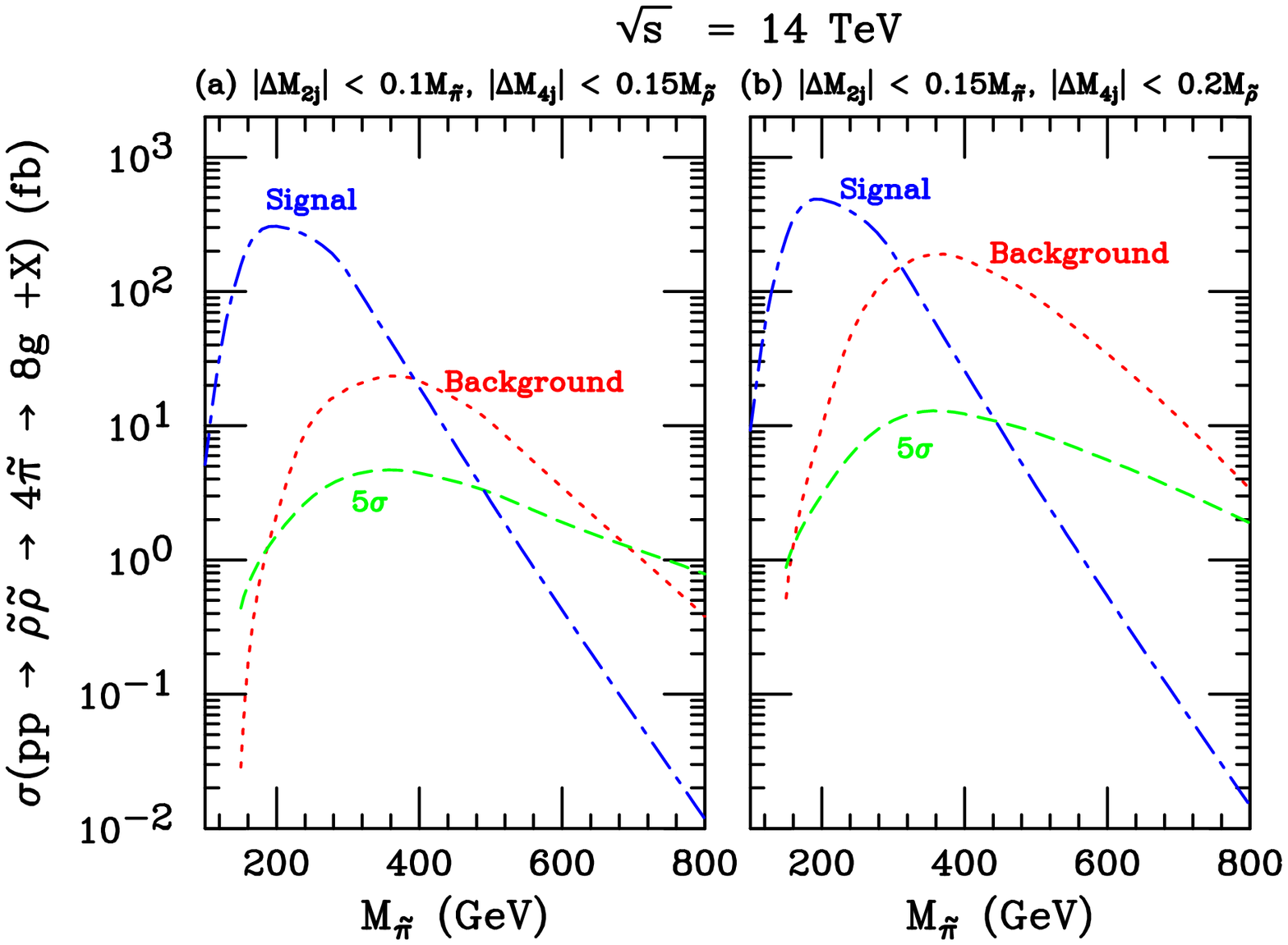}
\caption[]{
The cross section for $pp \to \trho\trho \to 4\tpi \to 8g +X$
(blue dot-dashed line) 
%at the LHC with $\sqrt{s} = 14$ TeV,
as a function of $M_{\tpi}$ with acceptance cuts on $p_T$, $\eta$,
and $\Del R$.
We have applied two sets of fixed mass cuts:
(a) $|M_{2j} -M_{\tpi}| < 0.10 M_{\tpi}$ and
    $|M_{4j} -M_{\trho}| < 0.15 M_{\trho}$,
or
(b) $|M_{2j} -M_{\tpi}| < 0.15 M_{\tpi}$ and
    $|M_{4j} -M_{\trho}| < 0.20 M_{\trho}$.
\Cblue{The $p_T$ cuts used were 
$p_T(j_1,\ldots,j_8)>320,250,200,160,125,90,60,40$ GeV.}
Also shown are the SM background cross section ($\sigma_b$)(red dotted line) 
and the minimal signal cross section that is required by a 5 sigma
criterion (green dashed line) with an integrated luminosity of 30 fb$^{-1}$.
\label{fig:fixed1}
}
\end{figure}

\begin{figure}[htb]
\centering\leavevmode
\epsfxsize=3.2in
\epsfbox{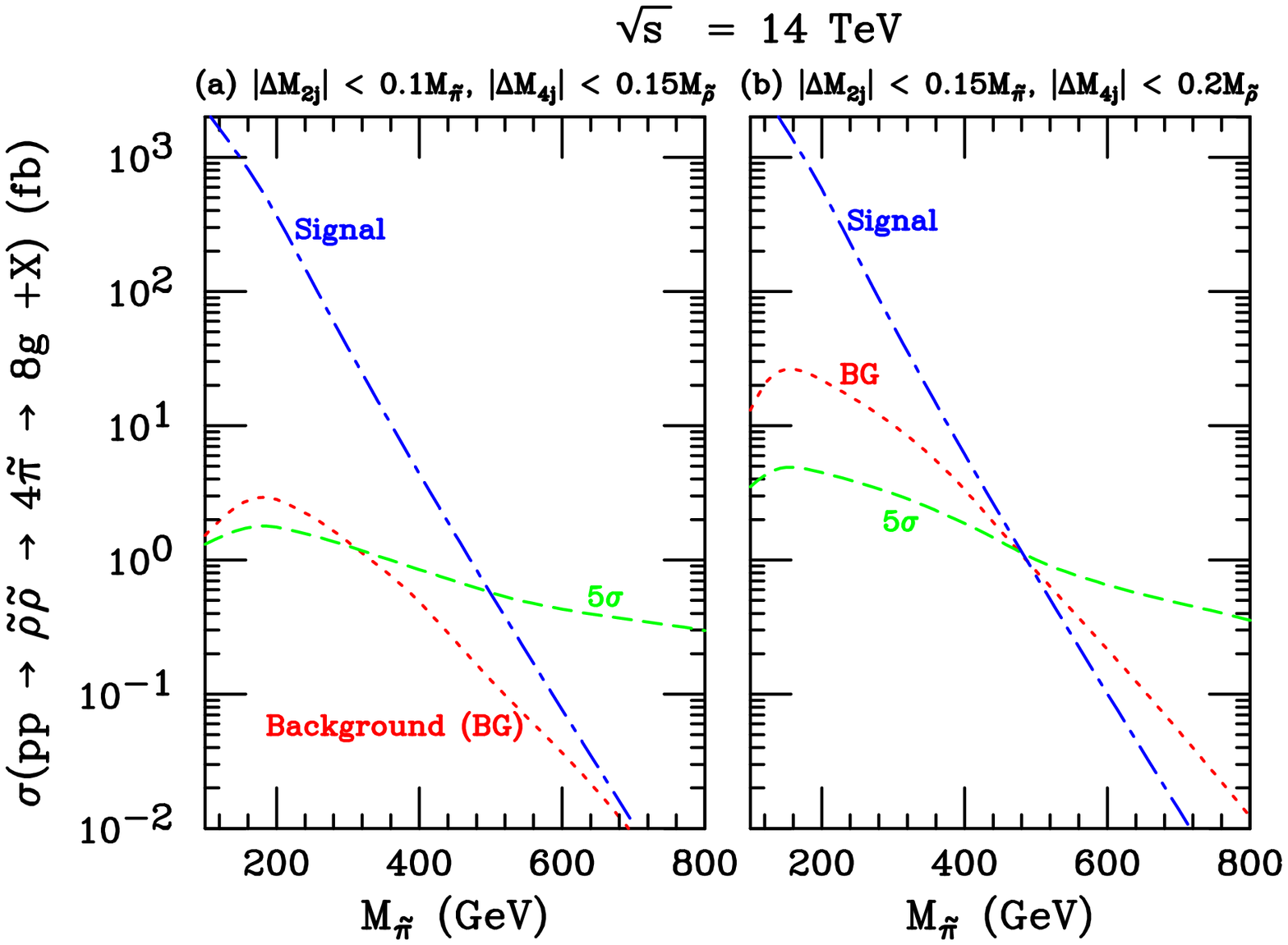}
\caption[]{
The cross section for $pp \to \trho\trho \to 4\tpi \to 8g +X$
(blue dot-dashed line) 
%at the LHC with $\sqrt{s} = 14$ TeV,
as a function of $M_{\tpi}$ with acceptance cuts on $p_T$, $\eta$,
and $\Del R$.
We have applied two sets of fixed mass cuts:
(a) $|M_{2j} -M_{\tpi}| < 0.10 M_{\tpi}$ and
    $|M_{4j} -M_{\trho}| < 0.15 M_{\trho}$,
or
(b) $|M_{2j} -M_{\tpi}| < 0.15 M_{\tpi}$ and
    $|M_{4j} -M_{\trho}| < 0.20 M_{\trho}$.
\Cblue{This figure differs from Fig.\ref{fig:fixed1} in that the $p_T$
  cuts used here are 
$p_T(j_1) > 1.5M_{\tpi}, \, p_T(j_2) > 1.2M_{\tpi}, \, p_T(j_3) > M_{\tpi}, 
\, p_T(j_4) > 0.8M_{\tpi}, \, p_T(j_5,j_6,j_7,j_8) > 50 \, {\rm GeV}$.}
Also shown are the SM background cross section ($\sigma_b$)(red dotted line) 
and the minimal signal cross section that is required by a 5 sigma
criterion (green dashed line) with an integrated luminosity of 30 fb$^{-1}$.
\label{fig:fixed2}
}
\end{figure}

Our results for the signal and background at the LHC with $\sqrt{s} = 14$ TeV 
are presented in Fig.~\ref{fig:relative}, \ref{fig:fixed1} and
\ref{fig:fixed2}.
To estimate the discovery potential at the LHC we include curves that 
correspond to the minimal cross section of signal ($\sigma_s$) 
required by our discovery criterion described in the following.
We define the signal to be observable
if the lower limit on the signal plus background is larger than
the corresponding upper limit on the background \cite{HGG} 
with statistical fluctuations
\begin{eqnarray}
L (\sigma_s+\sigma_b) - N\sqrt{ L(\sigma_s+\sigma_b) } \ge 
L \sigma_b +N \sqrt{ L\sigma_b }
\end{eqnarray}
%\Cblue{which corresponds to
\Cgreen{or equivalently,}
\begin{equation}
\sigma_s \ge \frac{N}{L}\left[N+2\sqrt{L\sigma_b}\right] \, .
\end{equation}
Here $L$ is the integrated luminosity, 
$\sigma_s$ is the cross section of the coloron signal,
and $\sigma_b$ is the background cross section.
The parameter $N$ specifies the level or probability of discovery. 
We take $N = 2.5$, which corresponds to a 5$\sigma$ signal.
For $\sigma_b \gg \sigma_s$, this requirement becomes similar to 
\begin{eqnarray*}
N_{\rm SS} = \frac{N_s}{\sqrt{N_b}}
 = \frac{L\sigma_s}{\sqrt{L\sigma_b}} \ge 5 \, ,
\end{eqnarray*}
where 
$N_s$ is the signal number of events, 
$N_b$ is the background number of events,
and $N_{\rm SS} =$ the statistical significance, which is 
commonly used in the literature. 
If the background has fewer than 25 events for a given luminosity, 
we employ the Poisson distribution and require that 
the Poisson probability for the SM background to fluctuate to this
level is less than $2.85\times 10^{-7}$.

%------------------------
% FIG 14
%------------------------

Fig.~\ref{fig:relative} shows the results for the {\it relative} mass
window scheme with 
$(\Delta M_{ij} = 30 \, {\rm GeV}, \Delta M_{4j} =60 \, {\rm GeV})$  
and 
$(\Delta M_{ij} = 50 \, {\rm GeV}, \Delta M_{4j} =100 \, {\rm GeV})$.
We have used the ordered $p_T$ cuts,
$p_T(j_1,\ldots,j_8)>320,250,200,160,125,90,60,40$ GeV. These are the 
momentum cuts used for the low mass example in Ref.~\cite{Kilic:2008ub}.
They are optimized for a $\tilde\pi$ mass around $225$ GeV.
With the cuts described above and an integrated luminosity of 30 fb$^{-1}$
we can look for detection of hyper-pion masses out to 440 GeV 
($M_{\tilde\rho} =$ \Cblue{1460 GeV}) 
with the $\Delta M_{ij}/\Delta M_{4j} = 50/100$ GeV window or 
out to $M_{\tilde{\pi}} =$ 400 GeV ($M_{\tilde{\rho}}=1333$ GeV) 
with the $30/60$ GeV windows.

%\Cred{ comment: Make a brief comment why $p_T$ cuts are chosen this way}

 Figs.~\ref{fig:fixed1} and \ref{fig:fixed2} show our results for
 the {\it fixed} mass window scheme. We include two choices of $p_T$ cuts
 and two mass window sizes as follows: 
(a)fixed mass windows with 
$\Del M_{ij} = 0.10 M_{\tpi}$ and $\Del M_{4j} = 0.15 M_{\trho}$, 
or (b)fixed mass windows with
$\Del M_{ij} = 0.15 M_{\tpi}$ and $\Del M_{4j} = 0.20 M_{\trho}$.
\Cblue{For Fig.~\ref{fig:fixed1} we apply the ordered $p_T$ cuts as
  used above for the relative mass cuts,
$p_T(j_1,\ldots,j_8)>320,250,200,160,125,90,60,40$ GeV. 
For Fig.~\ref{fig:fixed2} we use ordered cuts that scale as
the hyper-pion mass for the four leading jets, 
$p_T(j_1,\ldots,j_4) > 1.5 M_{\tpi},\,\, 1.2 M_{\tpi},\,\, 1.0 M_{\tpi},\,\, 
0.8 M_{\tilde\pi}$, and $ p_T(j) > 50$ GeV for the four lowest $p_T$ jets.}

The scaled $p_T$ cuts approximately capture the behaviour of the $p_T$ distribution peaks for the leading jets in the model. At high masses the
background could be further reduced by increasing the lower four $p_T$ thresholds with relatively small reduction of the signal. However, above
hyper-pion masses $\sim 450$ GeV  both the signal and the background are too small to make detection of either likely with
$30~ \text{fb}^{-1}$ of integrated luminosity. At low masses these could be lowered to capture more of the signal without
drastically enhancing the background. In general the $p_T$
cuts could be tailored to reduce the background below signal for the entire mass range shown, but we do not want to be overly reliant on the
model parameters we have chosen or to reduce the signal below practical detection limits. These cuts are a compromise to demonstrate the potential
discrimination of signal from background due to the boost of massive decaying particles.

The fixed mass windows scale as the masses to capture the similarly 
scaling width of the \Cblue{coloron} and the energy resolution of 
a real detector. Windows of $\Delta M_{ij} = 0.1 M_{\tilde\pi}$ and 
$\Delta M_{4j} = 0.15 M_{\tilde\rho}$ will require excellent resolution 
to be fully efficient. As in the relative window plots, we include 
a dashed line to estimate the discovery potential.

\section{Summary and Conclusions}

Colorons, massive vector bosons in the color-octet representation, 
are a generic possibility for exotic physics which could be detected 
at the LHC.
%Kilic et al. have shown how 
Both colorons and a set of scalar color-octet partners may emerge as the low
 energy phenomena of a generic new gauge group which
becomes confining at high energy scales. This has the interesting
consequence that relatively light colorons may evade dijet detection
bounds by decaying first to a pair of hyper-pions, each of which
decays into a pair of gluons. Based on analogy with the chiral
symmetry breaking interpretation of the Standard Model light mesons,
the parameters of the theory can be determined in terms of a single
unknown variable, $M_{\trho}$, at least for the case where
the new hyper-color gauge group is $SU(3)$. Further, the
phenomenology of the model should be fairly general even if we relax
these assumptions.

We have implemented this model in the MadGraph framework and checked 
the resulting code against analytical computations, finding good agreement. 
Using this we have analyzed the signal 
$pp \to \tilde\rho\tilde\rho \to 4\tilde\pi \to 8g +X$ at the LHC. 
In this channel we can reconstruct both the coloron and hyper-pion 
resonances better than in the single coloron channel by using 
correlations between invariant masses.

We have simulated both the signal and background using two mass cut 
schemes, a relative window scheme which requires no 
foreknowledge of the relevant masses, and a fixed window scheme 
which demonstrates the power to discriminate against the background 
for specific choices of candidate masses. 
We find that with $30\, \text{fb}^{-1}$ of integrated luminosity 
we can potentially detect such particles up to 
$M_{\tpi} \simeq$ 495 GeV and $M_{\trho} \simeq$ 1650 GeV. 
For other integrated luminosities the reach in $M_{\tilde{\rho}}$, 
for one choice of cuts, is shown in Table I.

%------------------------
% Table I
%------------------------

\begin{table}[htb]
\label{Invariant-Mass}
\caption[]{
Discovery reach in the coloron mass, using the fixed mass cuts of 
Fig.~15(a), for several values of integrated luminosity.
}
\begin{tabular}{|l|c|c|c|c|c|}
\hline
\bf{Integrated Luminosity}(fb$^{-1}$): & \bf{1} & \bf{10} & \bf{30} &
\bf{100} & \bf{1000} \\
\hline
\bf{Discovery Reach in $M_{\trho}$}(GeV): & \bf{1250} & \bf{1515} & \bf{1650} &
\bf{1780} & \bf{2080} \\
\hline
\end{tabular}
\end{table}

% Make next paragraph appear together.
\bigskip

In general, of course, we are sensitive to the choice of scales 
at tree level and a full predictive calculation would need to include 
effects of hadronization and jet reconstruction. 
Nonetheless, for a considerable range of parameters the signal can 
easily exceed the background even allowing for correction factors 
of order one. Our prospects improve significantly for increased jet 
energy resolution, leading to better reconstruction of invariant masses.

For the variety of related models with a similar phenomenology to the one 
we have considered, the fact that the hyper-mesons can be strongly 
produced once their mass thresholds are obtained, while avoiding 
the current dijet exclusion bounds, means they could be copiously 
produced at the LHC. As we gain experience with this collider, we have 
an excellent chance of discovering hyper-mesons or similar particles, 
if they exist, with masses up to approximately 2 TeV.

%------------------------------------------------------------------------------
% Acknowledgments
%------------------------------------------------------------------------------
\section*{Acknowledgments}

CK would like to thank the Kavli Institute of
Theoretical Physics at Santa Barbara for its hospitality.
This research was supported
in part by the U.S. Department of Energy
under Grants
No.~DE-FG02-04ER41305,
No.~DE-FG03-93ER40757,
No.~DE-FG02-04ER41306 and
No.~DE-FG02-04ER46140.

%------------------------------------------------
% NEW PAGE
%------------------------------------------------
\newpage

%-----------------------------------------------------------------------
% Bibliography
%-----------------------------------------------------------------------

%-----------------------------------------------------------------------
%   END DOCUMENT
%-----------------------------------------------------------------------
\end{document}